\documentclass{aa}
 \pdfoutput=1 
\usepackage[varg]{txfonts}

\usepackage{graphicx}

\usepackage{natbib}

\def\apjl{ApJL }

\def\apjs{ApJS }
\def\aap{A\&A }

\bibpunct{(}{)}{;}{a}{}{,} 

\begin{document} 

\title{A high-redshift quasar absorber without C\,{\sc iv} - a galactic outflow caught in the act?}

\authorrunning{A. Fox, P. Richter}

\author{Anne Fox\inst{\ref{inst1}} \and Philipp Richter\inst{\ref{inst1}}}

\institute{Institut für Physik und Astronomie, Karl-Liebknecht-Str. 24/25, 14476 Potsdam, Germany}\label{inst1} 

\date{Received 27/11/2015 / Accepted 15/02/2016}

\abstract {
We present a detailed analysis of a very unusual sub-damped Lyman $\alpha$ (sub-DLA) system at redshift $z=2.304$ towards the quasar Q\,$0453-423$, based on high signal-to-noise (S/N), high-resolution spectral data obtained with VLT/UVES. With a neutral hydrogen column density of $\log N(\text{H\,{\sc i}})=19.23$ and a metallicity of $-1.61$ as indicated by [O\,{\sc i}/H\,{\sc i}] the sub-DLA mimics the properties of many other optically thick absorbers at this redshift. A very unusual feature of this system is, however, the lack of any C\,{\sc iv} absorption at the redshift of the neutral hydrogen absorption, although the relevant spectral region is free of line blends and has very high S/N.     
Instead, we find high-ion absorption from C\,{\sc iv} and O\,{\sc vi} in another metal absorber at a velocity more than $220$\,km\,s$^{-1}$ redwards of the neutral gas component. We explore the physical conditions in the two different absorption systems using Cloudy
photoionisation models. We find that the weakly ionised absorber is dense and metal-poor while the highly ionised system is thin and more metal-rich. The absorber pair towards Q\,$0453-423$ mimics the expected features of a galactic outflow with highly ionised material that moves away with high radial velocities from a (proto)galactic gas disk in which star-formation takes place. We discuss our findings in the context of C\,{\sc iv} absorption line statistics at high redshift and compare our results to recent galactic-wind and outflow models.}


\maketitle

\section{Introduction}

High-redshift gas clouds in and around galaxies and in the intergalactic medium (IGM) give rise to absorption lines in the spectra of distant background quasars. Since many of these absorption line systems, in particular the strong ones, are thought to be associated with early galaxies, their study provides a useful means to investigate the progenitors of present-day galaxies, being unbiased with respect to distance, luminosity, and morphology of the host galaxies. 
In the canonical classification scheme, intervening absorbers are classified into four categories according to their neutral hydrogen column density. 
Gas clouds with neutral hydrogen column densities $N$(H\,{\sc i}) of less than $10^{17}$cm$^{-2}$ are called Lyman $\alpha$ forest systems \citep[see e.g.][]{Meiksin2009}. 
Up to $10^{19}$cm$^{-2}$, the absorbers are called Lyman limit systems (LLSs). 
Absorption line systems with neutral hydrogen column densities $>2 \times 10^{20}$cm$^{-2}$ are called damped Lyman $\alpha$ systems (DLAs) because of the pronounced damping wings that the hydrogen Lyman $\alpha$ absorption line exhibits at these high column densities. 
DLAs are assumed to be predominantly neutral because of the efficient self-shielding of the gas at such high column densities \citep{Schaye2001,Nagamine2010,Yajima2012,Rahmati2013}.
Finally, the sub-damped Lyman $\alpha$ systems (sub-DLAs) comprise the transition between Lyman limit systems and DLAs with column densities $10^{19}$cm$^{-2} \leq N$(H\,{\sc i})$\leq 2 \times 10^{20}$cm$^{-2}$ \citep{Dessauges2003,Peroux2003}. 

High-column density absorbers usually exhibit a rich spectrum of metal absorption lines from weakly and highly ionised species that can be used to explore the chemical composition as well as the physical conditions in the gas. 
Compared to other metal ions, absorption from three-times ionised carbon (C\,{\sc iv}) seems to be ubiquitous in intervening absorption line systems \citep{Fox2011}. 
It has been observed in each of these categories of absorber: in DLAs \citep{Rafelski2012,Fox2007b,Wolfe2005,Wolfe2000,Ellison2000,Lu1996}, in sub-DLAs \citep{Som2013,Fox2007a,Richter2005,Dessauges2003, Peroux2003} and in LLSs \citep{Prochter2010,Kirkman1999,Bergeron1994}. \citet{Fox2007a} even state that, in their sample of 63 DLAs and 11 sub-DLAs, no systems were found where C\,{\sc iv} absorption is not present. The overall presence of C\,{\sc iv} in QSO absorbers is usually  interpreted in terms of the presence of diffuse, metal-enriched gas that is photoionised (e.g. by the UV background) or collisionally ionised. In any case, the omnipresence of C\,{\sc iv} in QSO absorbers indicates that triply ionised carbon traces a gas phase that is characteristic for many astrophysical gaseous environments.

Highly ionised metal absorption lines far away from galaxies in the IGM could originate from metal enrichment by an early population of stars before the epoch of reionisation, as suggested by \citet{Cowie1998Nature,Gnedin1997} and \citet{Madau2001}. Studies of the evolution of the amount of C\,{\sc iv} over a large range in redshifts favour this scenario since it is consistent with only a mild or no evolution of the cosmological mass density of C$^{3+}$ ions \citep{Songaila2001,Songaila2005,Schaye2003,Boksenberg2003,Pettini2003,Ryan-Weber2006, Simcoe2006}. 
Also, \citet{Martin2010} study the size of the metal-enriched regions and argue that the metals must have been deposited by an earlier generation of galaxies. 
In this so-called early enrichment scenario, the O\,{\sc vi} and C\,{\sc iv} detections in DLAs indicate the presence of a hot ($\sim 10^{6}$\,K) gas phase surrounding the host galaxies. Such hot coronae are predicted to arise by shock heating of gas falling into potential wells \citep{ReesOstriker1977,Dave2001,Birnboim2003,Keres2005,Kang2005}. 
Studies by \citet{Kwak2010} and \citet{Fox2011} suggest a model in which the highly ionised species originate in turbulent mixing layers. 
In this scenario, these ions trace the radiatively cooling boundary layers between cool ($\sim 10^{4}$\,K) clouds and hot ($\sim 10^{6}$\,K) plasma. 
This model can explain the metallicity-independence of the strength of O\,{\sc vi} absorption in galaxy haloes \citep{Heckman2002,Fox2011a}, but it does not explain the high-ion kinematics which are generally more extended than the low-ion kinematics.
 
Alternatively, \citet{Adelberger2005} suggest that large-scale outflows of star-forming galaxies enriched the IGM with highly ionised metals at early times. 
The compelling evidence for mass outflows from high-redshift galaxies supports this picture \citep{Erb2006,Steidel2004,Pettini2001}. 
A discussion of observational evidence for galactic winds in nearby star-forming and active galaxies and in the high-redshift Universe can be found in \citet{Veilleux2005}. 
Next to observations, simulations of outflows predict the level of ionisation in these winds to be high \citep{Kawata2007,Fangano2007,Oppenheimer2006}. 
\citet{Fox2007a} discuss a scenario in which the star formation activity provides the metals as well as the ionisation energy that is needed to produce the hot ionised gas. 
Subsequent supernovae and stellar winds inject mechanical energy that causes the observed large line widths of the highly ionised species. 
Studies by \citet{D'Odorico2010,D'Odorico2013,Becker2011} and \citet{Ryan-Weber2009} support this scenario, because they show an increase in the C\,{\sc iv} cosmic mass density towards lower redshift and a downturn in the number density of C\,{\sc iv} systems towards $z=6$. Nevertheless, this is consistent with a decrease in overall metallicity towards higher redshifts.

Galactic-scale superwinds have important consequences for structure-formation \citep{Heckman2001}. 
Firstly, they are required to regulate star formation in almost all theoretical models of galaxy formation \citep[e.g.][]{Efstathiou2000}. 
Secondly, they enrich the surroundings of the galaxy and, if the metals remain in a hot phase, they may help to solve the problem of the missing metals at high redshift \citep{Sommer-Larsen2008,Bouche2005,Bouche2006,Bouche2007,Ferrara2005,Pettini1999,Pagel1999}.
Thirdly, the winds cause openings in the interstellar medium of the galaxies which allow Lyman continuum photons to escape and help to reionise the IGM \citep{Steidel2001}. 

In this overall context, studies of high-$z$ absorption line systems that exhibit an unusual behaviour with regard to the abundance and kinematics of highly-ionised species such as C\,{\sc iv} and O\,{\sc vi} represent important bench marks to test the various models and scenarios for diffuse ionised gas in the early Universe and to constrain them. 

We present here a detailed study of a sub-DLA at $z=2.304$ towards the quasar Q\,$0453-423$, which lacks absorption of C\,{\sc iv}, but that is accompanied by a second, redshifted absorber that possibly traces outflowing gas. 
The remainder of the paper is organised as follows. The data set is described in Section 2. We present the results of our analysis of the velocity structure and the ionisation modelling of the different gas phases in Section 3. We present possible scenarios for the observed absorption structure in Section 4, where we also discuss the implications of our findings. Finally, we summarise our work in Section 5. 

\section{Observations, data handling, and analysis method}

The data used in this study are publicly available in the UVES database within ESO's Science Archive Facility\footnote[1]{\texttt{http://archive.eso.org}}. They have been obtained in January 2002 as part of the ESO Large Programme 166.A-0106(A) (PI: J. Bergeron). Q\,0453$-$423 was observed through a 1\,arcsec slit and the total integration time was 98\,786\,s. The spectra were recorded at high spectral resolution ($R\sim45\,000$ or 6.6\,km s$^{-1}$ FWHM) and exhibit a very high S/N of up to $\sim100$ at $5000$\,\AA. The reduction of the raw data has been performed during the UVES Spectral Quasar Absorption Database programme
(SQUAD; PI: Michael T. Murphy) using a modified version of the UVES pipeline. The final combined spectrum covers a wavelength range of $3050-10\,430$\,\AA. 

In the metal absorption system at $z=2.304$ towards Q\,0453$-$423, we identify absorption lines arising from 9 different ions
(see Fig.\ \ref{fitovervelocityscale}), namely H\,{\sc i}, C\,{\sc ii}, O\,{\sc i}, Mg\,{\sc ii}, Al\,{\sc ii}, Si\,{\sc ii}, Si\,{\sc iii}, Si\,{\sc iv}, and Fe\,{\sc ii}. It is astonishing that we do not find any C\,{\sc iv} absorption within the same redshift range as the weakly ionised species, as the spectral region of the C\,{\sc iv} $\lambda 1548$ line is free from blends and the local S/N per resolution element is as high as 100. There is, however, a much weaker metal absorption system on the same line of sight with a velocity offset of more than 200\,km\,s$^{-1}$. In that system, we identify H\,{\sc i} absorption and weak C\,{\sc iii}, C\,{\sc iv}, and O\,{\sc vi} absorption, but no weakly ionised species and no Si\,{\sc iv} (as discussed in detail in Sect.\,3).
 
We have analysed the UVES spectrum using the \texttt{FITLYMAN} package implemented in MIDAS \citep{Fontana1995}. Wherever possible, we have fitted absorption lines with coincident redshifts simultaneously. In addition, we have fitted multiple absorption lines of the same ion as multiplet to ensure they have the same column density. For weak or undetected absorption features, we determine the limiting equivalent width based on the local S/N and considering a 3 $\sigma$ detection limit for unsaturated line absorption. This value serves as upper limit except for C\,{\sc ii}. Since the C\,{\sc ii} absorption is blended, we have derived upper and lower limits by adjusting the continuum in a way that it either includes or excludes all of the blend.

In our analysis, we include statistical errors, which are given by \texttt{FITLYMAN}. We do not account for contributions from errors in continuum placement, systematic uncertainties from fixing of the line centres and $b$-values, and uncertainties in the atomic data. Throughout the paper we use solar abundances from \citet{Asplund2009} and atomic data from \citet{Morton2003}.

\section{Results}

\subsection{Velocity structure}



\begin{figure*}
 \centering
\includegraphics[width=16.5cm]{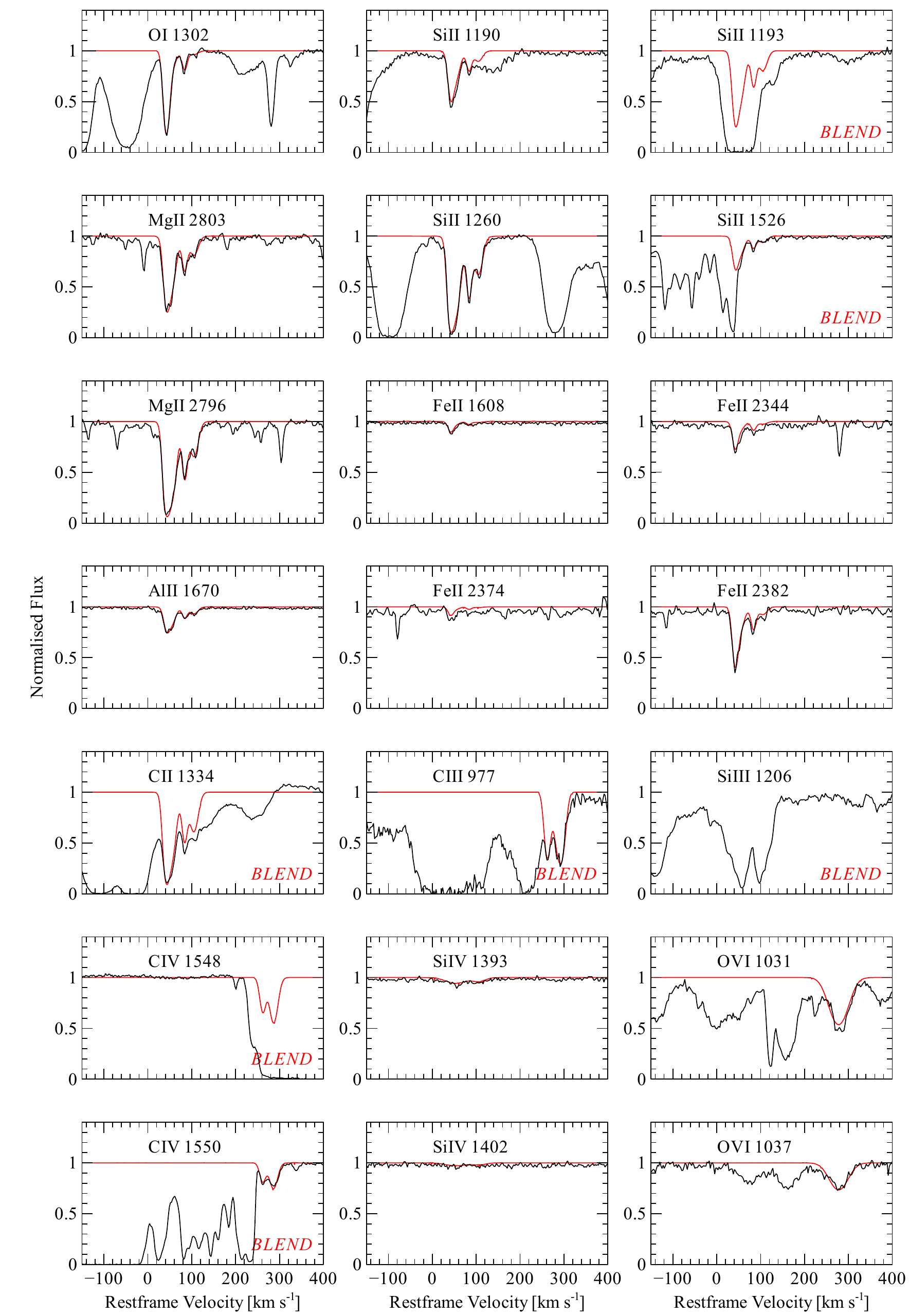} 
\caption{Stacked velocity plot for the absorption system at $z=2.304$ toward Q\,0453$-$423. The normalised flux is represented by the black line. Where lines have been fitted, the fit is represented by the red line. Parameters of the fit can be found in Table \ref{Fit results}. The figure also indicates which lines are blended.}
\label{fitovervelocityscale}
\end{figure*}


Fig.\ \ref{fitovervelocityscale} shows absorption lines of all detected ions in this system. In the range of 0 -- 100\,km\,s$^{-1}$, we find the centre of the strong neutral hydrogen absorption as well as absorption lines of a variety of weakly ionised species: C\,{\sc ii}, O\,{\sc i}, Mg\,{\sc ii}, Al\,{\sc ii}, Si\,{\sc ii}, and Fe\,{\sc ii}. Strikingly, there is no high-ion absorption detected in this velocity interval except for a small amount of Si\,{\sc iv}. The C\,{\sc iv} $\lambda$1550 line is blended. Nevertheless, from the C\,{\sc iv} $\lambda$1548 panel (6th row in the left column of Fig.\ \ref{fitovervelocityscale}), it is obvious that there is no C\,{\sc iv} absorption in the same velocity range as the weakly ionised species are found down to limiting column density of $10^{12.1}$\,cm$^{-2}$.  However, we find absorption of C\,{\sc iii}, C\,{\sc iv}, and O\,{\sc vi} well separated from the weakly ionised species by more than 220\,km\,s$^{-1}$ ($z_{\rm abs}=2.307$). There is also neutral hydrogen absorption at this velocity, but no absorption from weakly ionised species.



\begin{figure*}
\includegraphics[width=17cm]{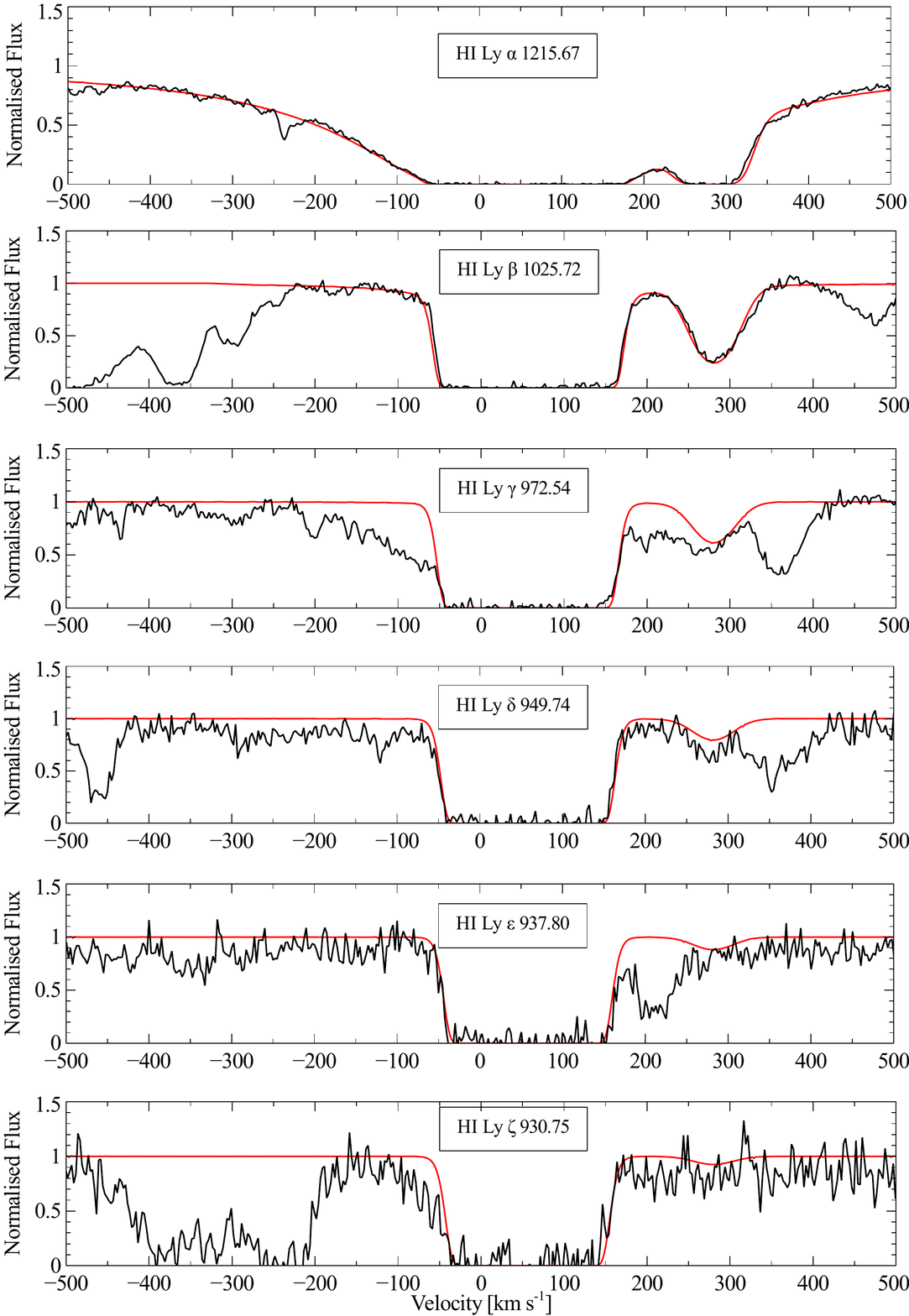}
\caption{Lyman-series absorption from neutral hydrogen in the sub-DLA at $z=2.304$ towards the quasar Q\,0453-423}
\label{hydrogenabsorption}
\end{figure*}


Fig.\ \ref{hydrogenabsorption} shows the Lyman-series absorption from neutral hydrogen absorption for the $z=2.304$ absorber, which consists of two components. The multi-component Voigt-profile fit of the absorption lines Lyman $\alpha$, $\beta$, $\gamma$, $\delta$, $\epsilon$, and $\zeta$ yields column densities of $\log N(\text{H\,{\sc i}})=19.23 \pm 0.01$ and $\log N(\text{H\,{\sc i}})=14.58 \pm 0.01$ for the blue and the red component, respectively (Table\ \ref{Fit results}). 
This classifies the $z=2.304$ absorber as sub-damped Lyman $\alpha$ system.




\subsection{Weakly ionised absorber}


For the metal ions, we identify four velocity sub-components at 42\,km\,s$^{-1}$, 52\,km\,s$^{-1}$, 84\,km\,s$^{-1}$, and 105\,km\,s$^{-1}$, i.e. they span a relatively narrow range of 63\,km\,s$^{-1}$. The two components at 84 and 105\,km\,s$^{-1}$ are substantially weaker than the other two components.



\begin{table}[htbp]
\centering
\caption{Component structure, column densities, and Doppler parameters for the weakly ionised phase of the $z=2.304$ absorber towards Q\,0453-423. \label{Fit results}}
\begin{tabular}{ccccc} \hline\hline
\noalign{\vspace{1mm}}
No&$v_{\text{rel}}$ [km s$^{-1}$]& Ion A & $\log N$(A) & $b$ [km s$^{-1}$]\\
\noalign{\vspace{1mm}}
\hline
\noalign{\vspace{1mm}}
1&42	&	C\,{\sc ii}&	 $13.73\pm0.07$	&	$7.5\pm0.3$	\\ 
&	&	O\,{\sc i}&	 $14.17\pm0.01$	&		\\
&	&	Mg\,{\sc ii}&	 $12.75\pm0.01$	&		\\
&	&	Al\,{\sc ii}&	 $11.48\pm0.07$	&		\\
&	&	Si\,{\sc ii}&	$12.88\pm0.04$	&	\\
&	&	Fe\,{\sc ii}&	 $12.75\pm0.01$	&		\\
2&52	&	C\,{\sc ii}&	 $13.83\pm0.04$	&	$13.5\pm0.5$	\\
&	&	O\,{\sc i}&	 $13.53\pm0.06$	&		\\
&	&	Mg\,{\sc ii}&	 $12.95\pm0.01$	&		\\
&	&	Al\,{\sc ii}&	 $11.83\pm0.03$	&		\\
&	&	Si\,{\sc ii}&	$12.99\pm0.03$	&	\\
&	&	Fe\,{\sc ii}&	 $12.51\pm0.02$	&		\\
3&84	&	C\,{\sc ii}&	 $13.32\pm0.03$	&	$7.2\pm0.3$	\\
&	&	O\,{\sc i}&	 $13.26\pm0.03$	&		\\
&	&	Mg\,{\sc ii}&	 $12.42\pm0.01$	&		\\
&	&	Al\,{\sc ii}&	 $11.32\pm0.02$	&		\\
&	&	Fe\,{\sc ii}&	 $12.26\pm0.02$	&		\\
&	&	Si\,{\sc ii}&	$12.53\pm0.01$	&	\\
4&105	&	C\,{\sc ii}&	 $13.40\pm0.02$	&	$12.4\pm0.5$	\\
&	&	O\,{\sc i}&	 $12.73\pm0.10$	&		\\
&	&	Mg\,{\sc ii}&	 $12.33\pm0.01$	&		\\
&	&	Al\,{\sc ii}&	 $11.33\pm0.03$	&		\\
&	&	Si\,{\sc ii}&	$12.44\pm0.01$	&	\\
&	&	Fe\,{\sc ii}&	 $11.96\pm0.05$	&		\\ \hline 
\noalign{\vspace{1mm}}
1&54	&	Si\,{\sc iv}&	 $12.25\pm0.05$	&	$31.8\pm4.5$	\\ 
&107	&	Si\,{\sc iv}&	 $11.86\pm0.08$	&	$19.1\pm3.1$	\\ 
\noalign{detection limit for Si\,{\sc iv} is 11.8} 
\noalign{\vspace{1mm}}\hline 
\noalign{\vspace{1mm}}
&&C\,{\sc iv} & $\leq12.1$ & \\
&&Al\,{\sc iii} & $\leq11.9$ & \\ \hline 
\noalign{\vspace{1mm}}
1&57	&	H\,{\sc i}&	 $19.23\pm0.00$	&	$34.4\pm0.1$	\\ 
2&281	&	H\,{\sc i}&	 $14.58\pm0.01$	&	$32.9\pm0.3$	\\ \hline
\noalign{\vspace{1mm}}
\noalign{\vspace{1mm}}
\end{tabular}
\tablefoot{ The weakly ionised metal species have been fit simultaneously. Additionally, Mg\,{\sc ii} and Fe\,{\sc ii} have each been fit as a multiplet with fixed line centres from the simultaneous fitting. For hydrogen, Lyman $\alpha$, $\beta$, $\gamma$, $\delta$, $\epsilon$, and $\zeta$ have been fit simultaneously. For silicon, only the Si\,{\sc iv} $\lambda 1393$ line has been fit, since the line is very weak but the column density is above the detection limit. The errors are provided by \texttt{FITLYMAN}. 
}
\end{table}


Only one high ion is seen in absorption at the same redshift as the weakly ionised species. It can be seen in Fig.\ \ref{fitovervelocityscale} that the Si\,{\sc iv} absorption is very weak and the velocity structure is slightly different from that of the weakly ionised species. 



\begin{figure}
\resizebox{\hsize}{!}{\includegraphics{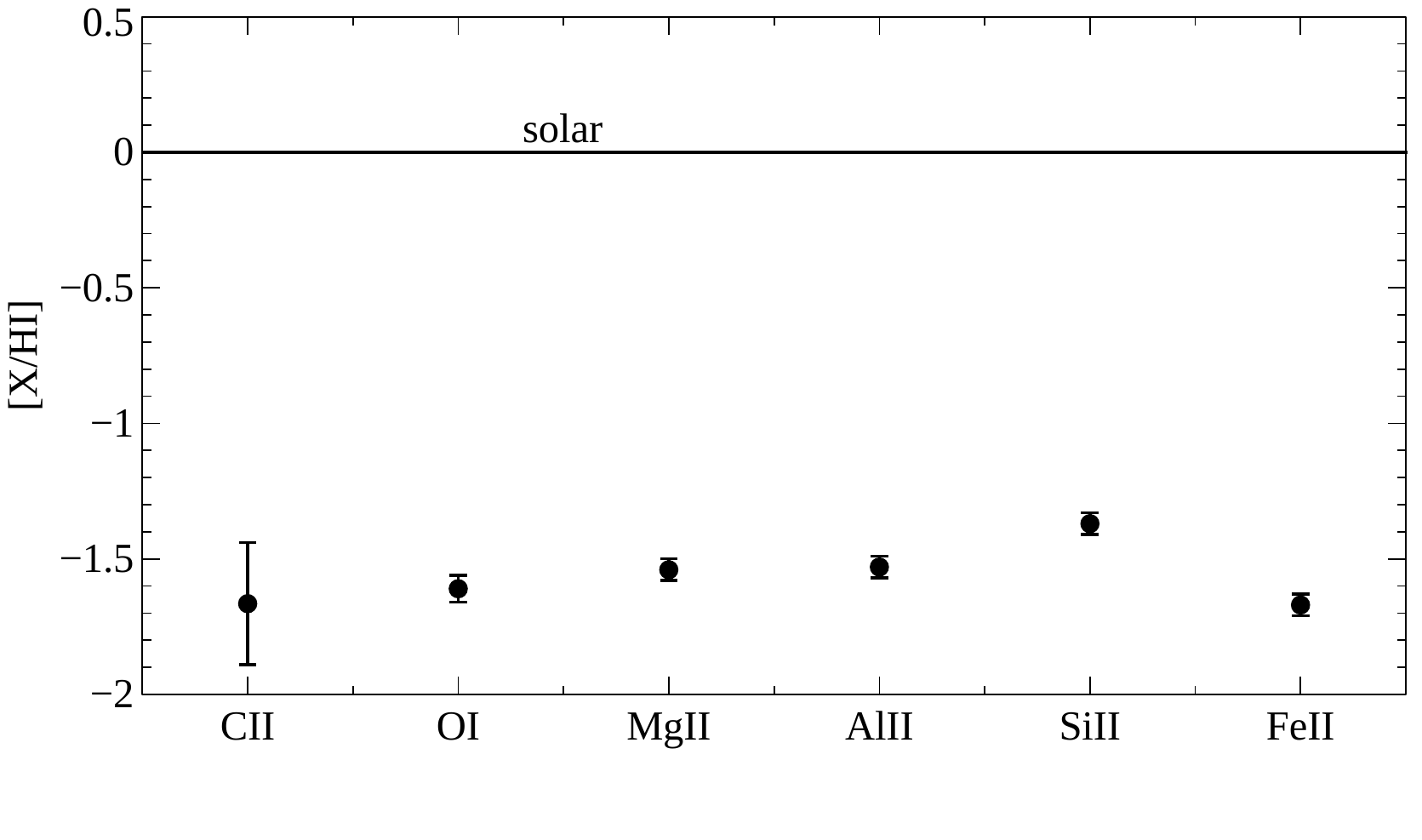}}
\caption{Abundances with respect to solar abundances in the sub-DLA at $z=2.304$ towards the quasar Q\,0453-423}
\label{abundances}
\end{figure}


Ion column densities and $b$-values for the individual components in the weakly-ionised absorber are presented in Table\ \ref{Fit results}. The table also shows upper limits for C\,{\sc iv}, Si\,{\sc iv}, and Al\,{\sc iii}, which correspond to the 3 $\sigma$ detection limits.

Fig.\ \ref{abundances} shows the metal abundances of the absorption system, estimated from the weakly ionised metal species and averaged over all velocity sub-components (see above) without taking into account any ionisation corrections (see also Table\ \ref{Element abundances table}).
The ionisation conditions in the gas will be discussed in Sect.\,4.  
The metallicity pattern seems fairly smooth except for a slight overabundance of silicon and a slight underabundance of oxygen, but these deviations are not as pronounced as in the system previously analysed by us in \citet{meinPaper2014}.


\begin{table*}
 \caption{Total column densities and elemental abundances for the weakly ionised absorber.}
 \centering
 \label{Element abundances table}
 \begin{tabular}{lllllll}\hline\hline
\noalign{\vspace{1mm}}
Ion A & $\log N$(A)$_{\text{tot}}$ & Element X& $\log N$(X)$_{\odot}$&$\log$(A/H\,{\sc i}) & [A/H\,{\sc i}] \\ 
\noalign{\vspace{1mm}}
\hline
\noalign{\vspace{1mm}}
C\,{\sc ii} & $13.77 \ldots 14.20^{1}$ 	& C &$8.43 \pm 0.05$ &$ -5.46 \ldots -5.03^{1}$ & $-1.89 \ldots -1.44^{1} $	\\
O\,{\sc i} & $14.31 \pm 0.01$ & O & $ 8.69 \pm 0.05 $ & $ -4.92 \pm 0.01 $ & $-1.61\pm 0.05$	\\ 
Mg\,{\sc ii} & $13.29 \pm 0.01$ & Mg& $ 7.60 \pm 0.04 $ & $ -5.94 \pm 0.01 $ & $-1.54\pm 0.04$	\\
Al\,{\sc ii} & $12.15 \pm 0.02$ & Al& $ 6.45 \pm 0.03 $ & $ -7.08 \pm 0.02 $ & $-1.53\pm 0.04$	\\
Si\,{\sc ii} & $13.37 \pm 0.02$ & Si& $ 7.51 \pm 0.03 $ & $ -5.86 \pm 0.02 $ & $-1.37\pm 0.04$	\\
Fe\,{\sc ii} & $13.06 \pm 0.01$ & Fe& $ 7.50 \pm 0.04 $ & $ -6.17 \pm 0.01 $ & $-1.67\pm 0.04$	\\
Si\,{\sc iv} & $12.40 \pm 0.04$ & & & & \\
\hline
\end{tabular}
\tablefoot{$^{1}$ The column density range takes into account the influence of the blend.} 
\end{table*}


\subsection{Highly ionised absorber}

Fig.\ \ref{fitovervelocityscale} also shows the absorption profiles of the highly ionised species C\,{\sc iii}, C\,{\sc iv}, and O\,{\sc vi}. A large velocity offset of $\gtrsim 200$ km s$^{-1}$ is obvious. Furthermore, the narrow velocity range ($\lesssim 40$ km s$^{-1}$) of the highly ionised species is striking. Previous studies of high ion absorption in DLAs \citep{Lu1996,Ledoux1998,Wolfe2000,Fox2011} and sub-DLAs \citep{Dessauges2003,Peroux2003,Richter2005} have found that the high ion profiles generally occupy an overlapping but more extended velocity range than the low ion profiles. This is also evident from Fig.\ 2 in \citet{Fox2007a}, where most of the systems have a C\,{\sc iv} total line width $\leq 500$\,km\,s$^{-1}$ and an absolute velocity offset of $\leq 150$\,km\,s$^{-1}$.

Also, it is evident from Fig.\ \ref{fitovervelocityscale} that O\,{\sc vi} shows a different absorption pattern than C\,{\sc iv}. This distinct behaviour of   O\,{\sc vi} has been found in previous studies \citep[e.g.][]{Lehner2014}. Therefore, we decided to determine the O\,{\sc vi} column density independently from the carbon fitting. O\,{\sc vi} has been fit as a multiplet and carbon has been fit under the assumption that the line centres of C\,{\sc iii} and  C\,{\sc iv} have the same $z$. The fit is also shown in Fig.\ \ref{fitovervelocityscale}. 
We identify three velocity components at 262\,km\,s$^{-1}$, 286\,km\,s$^{-1}$, 296\,km\,s$^{-1}$, i.e. the velocity offset to the bluest component of the weakly ionised species is 220\,km\,s$^{-1}$. Measured column densities and $b$-values for the highly ionised species are listed in Table \ref{highiontable}. The total column densities and upper limits of the high ion lines can be found in Table\ \ref{totalcoldenshigh}.
To verify our results, we have additionally measured the equivalent width of the O\,{\sc vi} absorption line and, under the assumption of an optically thin line, we have deduced from that a total column density of $\log N$(O\,{\sc vi})$= 13.80$, which is slightly lower than what has been determined by Voigt profile fitting. 


\begin{table}[htbp]
\centering
\caption{Component structure, column densities, and Doppler parameters for the highly ionised absorber towards Q\,0453-423. \label{highiontable}}
\begin{tabular}{ccccc} \hline\hline
\noalign{\vspace{1mm}}
No&$v_{\text{rel}}$ [km s$^{-1}$]& Ion A & $\log N$(A) & $b$ [km s$^{-1}$]\\
\noalign{\vspace{1mm}}
\hline
\noalign{\vspace{1mm}}
1&262	&	C\,{\sc iii}&	 $12.95\pm0.02$	&	$8.5\pm0.6$	\\ 
&	&	C\,{\sc iv}&	$12.95\pm0.01$	&	$8.5\pm0.4$	\\
2&286	&	C\,{\sc iii}&	 $12.91\pm0.04$	&	$11.2\pm0.8$	\\ 
&	&	C\,{\sc iv}&	$13.19\pm0.01$	&	$13.6\pm0.6$	\\
3&296	&	C\,{\sc iii}&	 $12.85\pm0.04$	&	$14.0\pm1.2$	\\ 
&	&	C\,{\sc iv}&	$12.10\pm1.24$	&	$1.0\pm0.7$	\\\hline 
\noalign{\vspace{1mm}}
1&278	&	O\,{\sc vi}&	 $13.94\pm0.01$	&	$28.4\pm0.7$	\\ \hline 
\noalign{\vspace{1mm}}
\end{tabular}
\tablefoot{C\,{\sc iii} and C\,{\sc iv} have been fit assuming that these two ions have the same velocity sub-components. The errors are provided by \texttt{FITLYMAN}. 
}
\end{table}

\begin{table}[htbp]
\centering
\caption{Total column densities for the highly ionised phase.\label{totalcoldenshigh}}
\begin{tabular}{ccccc} \hline\hline
\noalign{\vspace{1mm}}
Ion A & total $\log N$(A) \\
\noalign{\vspace{1mm}}
\hline
\noalign{\vspace{1mm}}
	C\,{\sc iii}&	 $13.38\pm0.02$		\\ 
	C\,{\sc iv}&	 $13.41\pm0.07$		\\
	O\,{\sc vi}&	 $13.94\pm0.04$		\\
	Si\,{\sc iii}&	 $<11.7$		\\
	Si\,{\sc iv}&	 $<11.8$		\\\hline
\noalign{\vspace{1mm}}
\end{tabular}
\tablefoot{The errors are provided by \texttt{FITLYMAN}. The upper limit given for Si\,{\sc iii} and Si\,{\sc iv} is the 3 $\sigma$ detection limit.
}
\end{table}







\section{Cloudy modelling}

\subsection{Model setup}

In order to test how much C\,{\sc iv} we can expect in a gas characterised by this pattern of weakly ionised metals, we have carried out photoionisation models using version 13.03 of the Cloudy photoionisation code \citep{Ferland1998}. We have assumed the incident ionising radiation field to be composed of the cosmic
background radiation and a Haardt \& Madau (2012) UV background \citep{Haardt2012}, both evaluated at the redshift of the absorber and scaled to $\log F(\nu)=-20.3608$ at 1.0\,Ryd. Here, $F(\nu) = 4 \pi J(\nu)$, where $J(\nu)$ is the mean intensity of the incident radiation field per unit solid angle, given in erg\,s$^{-1}$\,Hz$^{-1}$\,cm$^{-2}$.
We modelled a plane-parallel slab of photoionised gas and varied the hydrogen density $n_{\text{H}}$ between $10^{-5.0}$ and $10^{-0.4}$\,cm$^{-3}$ in steps of $10^{-0.2}$ cm$^{-3}$. The temperature was left as a free parameter. 

The simulations were stopped once the observed total H\,{\sc i} column density was reached. We scaled down the overall metallicity in the model to 2.5\% solar which corresponds to the measured [O\,{\sc i}/H\,{\sc i}]. This value has been adopted as starting value for the models' metallicity because of the strong connection between O\,{\sc i} and H\,{\sc i} through charge-exchange reactions. For our final model, we chose the hydrogen density and metallicity that reproduce the observations best, i.e., that are consistent with the column densities determined from the observations. 
For this absorption system, we have considered the total column densities and have not tried to model each component individually since the metallicity pattern is similar in the individual components. 
However, we have set up a separate model for the highly ionised gas with the significant velocity offset.


\begin{figure*}
 \noindent
\begin{minipage}[t]{.46\linewidth}
\includegraphics[width=\linewidth]{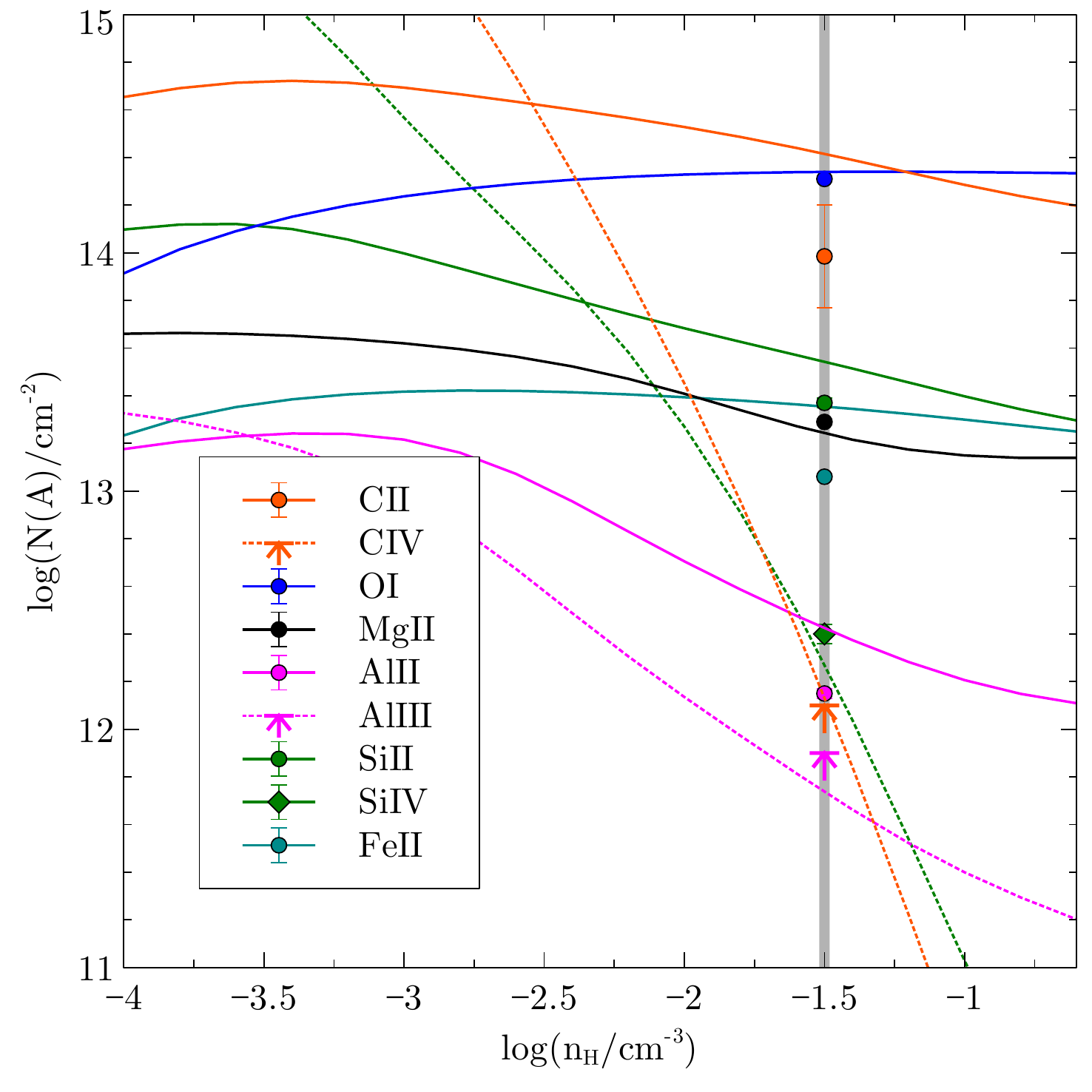}  
a) Metallicity of $2.5\%$ solar.
\end{minipage} \hfill
\begin{minipage}[t]{.46\linewidth}
 \includegraphics[width=\linewidth]{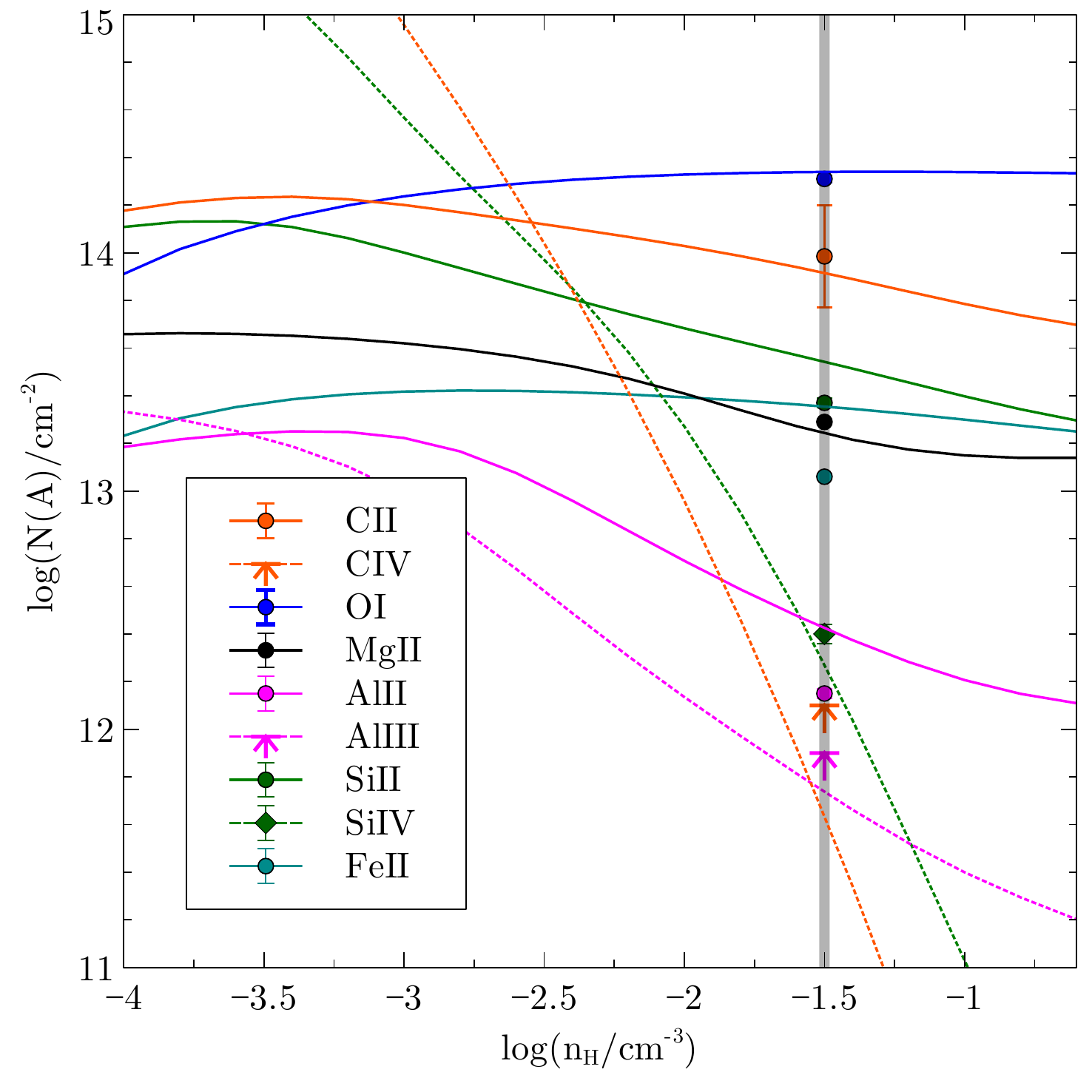}  
b) Metallicity of $2.5\%$ solar, but (C/H)=$-4.07$ which corresponds to [C/O]$=-0.5$.
\end{minipage} 
\caption {Cloudy models for the weakly ionised metal species in the absorption line system at z = 2.304. The lines represent the Cloudy predictions for the column densities versus the gas density $\log n_{\text{H}}$, with the observations (symbols) plotted at best-fit value of $\log n_{\text{H}}$ as indicated by the vertical grey line. The correlation between the symbols and the lines is shown in the keys in the lower left corners. The error bars are in most cases smaller than the symbol itself. For C\,{\sc iv} and Al\,{\sc iii} upper limits are indicated as explained in the text.} \label{Cloudycoldphase}
\end{figure*}


\subsection{Weakly ionised absorber}\label{Cloudylow}

Fig.\ \ref{Cloudycoldphase} shows the Cloudy models for the weakly ionised gas phase based on the initial metallicity assumption of 2.5\% solar metallicity.
Upper limits are given for Al\,{\sc iii} and C\,{\sc iv}. They correspond to the detection limit.  
Since the C\,{\sc ii} line is blended, we give an estimate of the maximum and the minimum amount of C\,{\sc ii} in the system, including or excluding the blend, respectively. In either case, it is obvious that Cloudy predicts more C\,{\sc ii} than we observe. Therefore, we have deciced to reduce the carbon abundance to correspond to [C/O]$=-0.5$, as is commonly observed in DLAs \citep{Fox2007c,Fox2011,Cooke2011,Pettini2008} and in low-metallicity stars in the Galactic halo \citep{Akerman2004}. This model is represented in Fig.\ \ref{Cloudycoldphase}b). Now, the observed carbon and oxygen column densities match the Cloudy predicitions. The remaining ion species show minor discrepancies that can have various reason as we discussed in \citet{meinPaper2014}. In any case, it is obvious that the hydrogen density of this gas cloud must be relatively high, $n_{\rm H}\gtrsim 10^{-1.5}$ cm$^{-3}$. This density limit is comparable to the densities found in warm neutral gas in the disk and halo of the Milky Way \citep{Richter2009}. The total hydrogen column density in the cloud is calculated to log $N$(H$)=19.94$, from which we derive an absorber size of $d=N($H$)/n_{\rm H}= 892$\,pc. At the chosen hydrogen density, the temperature in the gas is constrained to $T \approx 1.3 \times 10^{4}$\,K.


\begin{figure}
 \resizebox{\hsize}{!}{\includegraphics{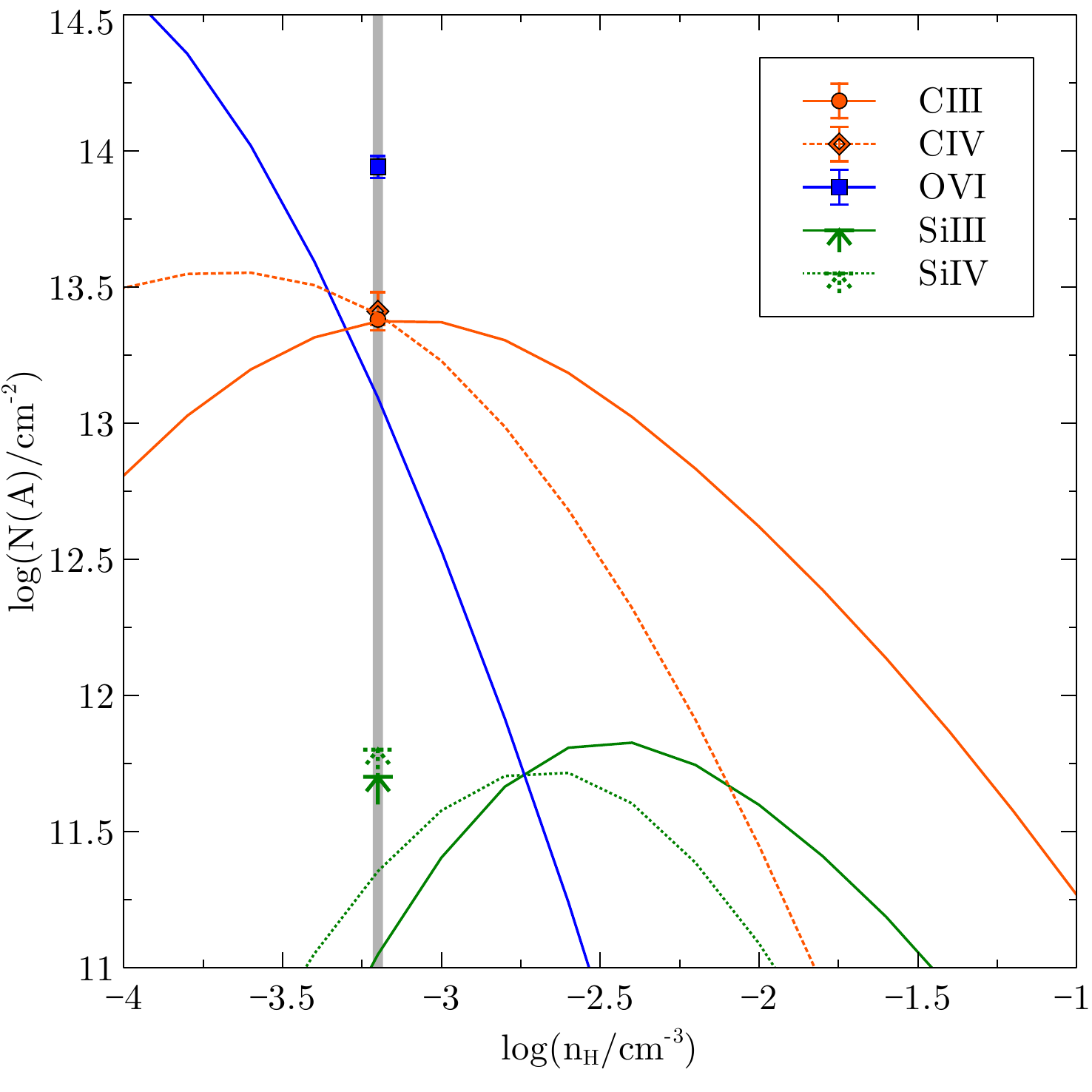}}
\caption {Cloudy model for the highly ionised metal species in the absorption line system at z = 2.304. The lines represent the Cloudy predictions
and the symbols the measured values. The correlation between the symbols and the lines is shown in the key in the upper right corner. For Si\,{\sc iii} and Si\,{\sc iv} upper limits are indicated as explained in the text. The error on the C\,{\sc iii} column density is propably larger than indicated because of possible line blending. The vertical grey line indicates the hydrogen density that causes the smallest discrepancies between the Cloudy predictions and the measured values. 
} \label{Cloudyhotphase}
\end{figure}


\subsection{Highly ionised gas absorber}

Fig.\ \ref{Cloudyhotphase} shows the Cloudy model for the highly ionised gas phase. For this model, the initial metallicity assumption had to be varied and increased up to 17\% solar metallicity. The simulations were stopped at a hydrogen column density of $\log N$(H\,{\sc i})$=14.58$, which has been measured for the redward hydrogen component as seen in Fig.\ \ref{hydrogenabsorption}. The total hydrogen column density is log $N$(H$)=18.59$, while the best matching density for this absorber is ($n_{\text{H}} \approx 10^{-3.2}$\,cm$^{-3}$). These numbers indicate an absorber size of $d=2.0$\,kpc. In Fig.\ \ref{Cloudyhotphase}, upper limits are given for Si\,{\sc iii} and Si\,{\sc iv}. The C\,{\sc iii} line is most likely blended, so the errors can be larger than indicated by the \texttt{FITLYMAN} fits. 

O\,{\sc vi} does not match the Cloudy model. In fact, it has often been found that O\,{\sc vi} does not reside in the same gas phase as C\,{\sc iv} and Si\,{\sc iv} \citep{Fox2007b,Fox2011,Lehner2008}. 
The column density of O\,{\sc vi} can be explained by the presence of a shock-heated gas phase with a temperature of $T=1.5 \times 10^{6}$\,K, where only a minimum amount of hydrogen is substracted from the C\,{\sc iv} gas phase. 

From our modelling, we find that the metallicity of the highly ionised phase is not the same as in the weakly ionised phase. It is generally higher and a carbon-oxygen-ratio of $-0.5$ does not match the observations in the highly ionised absorber.

\section{Discussion} 

\subsection{Constraints on physical conditions}

The lack of significant C\,{\sc iv} absorption in the $z=2.304$ 
sub-DLA towards Q\,$0453-423$ is a very unusual feature for
an intervening absorption system at this large gas-column density.
This aspect thus requires further consideration.

It is the intrinsic nature of basically all extended interstellar and 
intergalactic gas structures that the they exhibit an intrinsic distribution of gas densities that may span 
several orders of magnitudes. As a result of ionisation and 
recombination processes, such density variations
result in the formation of spatially distinct gas phases 
that coexist within the same superordinate structure. 
Abrupt changes in the gas density are not expected for a cosmic 
gaseous structure being in a steady state because of mixing
processes in the boundary layers between high-density and low-density 
regions. In fact, it has been suggested 
that a considerable fraction of intervening high-ion absorbers 
trace exactly such mixing layers \citep{Fox2011}, indicating 
the widespread presence of multi-phase intergalactic 
and circumgalactic gas at low and high redshifts.  

The absorber at $z=2.304$ towards Q\,$0453-423$
obviously lacks such a boundary layer, while highly-ionised gas is
redshifted by more than $220$\,km\,s$^{-1}$. This clearly indicates
that the gas very recently must have experienced some kind 
of dynamic event that has led to the kinematic separation 
of the different gas phases. 

With the density and temperature constraints of from the Cloudy modelling, the thermal pressure in the weakly ionised component is restricted to  
$P/k\gtrsim 387$\,cm$^{-3}$\,K, where $k$ is Boltzmann's constant.
Assuming that the gas is in pressure equilibrium (confined by gravity and/or by an ambient medium), this value suggests that the absorber is located deep within a potential well of a galaxy-size dark-matter halo. 

If the gas would be over-pressured for some reasons, however, it would expand to reach such pressure equilibrium. 
The relevant time scale is the free expansion time scale, $t_{\rm exp}$, which is defined by the ratio between absorber size, $d$, and the sound speed in the gas, $c_{\rm s}$. 
As shown by \citet{Schaye2007}, $t_{\rm exp}$
can be parametrised for intergalactic conditions in the form

\begin{equation}
t_{\rm exp}=\frac{d}{c_{\rm s}}\approx 6.4\times 10^6\,{\rm yr}\,
\left(\frac{d}{100\,{\rm pc}}\right)\,\left(\frac{T}{10^4\,{\rm K}}
\right)^{-1/2}.
\end{equation}

In our case, $d=892$ pc, so that 
with $T=1.3\times 10^4$ K we obtain $t_{\rm exp}\approx 4.9\times
10^7$ yr, which is very small compared to the Hubble time.
However, if the structure would expand within an ambient medium with
a smooth pressure gradient, it would be destroyed on a 
similarly short timescale due to Rayleigh-Taylor and Kelvin-Helmholtz
instabilities \citep{Schaye2007}. Therefore, such an object
would be very short-lived. A similar problem has been reported by \citet{Crighton2015}.

With these physical boundary conditions in mind, we explore subsequently possible scenarios for the origin of the
absorber at $z=2.304$ towards Q\,$0453-423$.


\begin{figure}
\begin{center}
\resizebox{0.6\hsize}{!}{\includegraphics{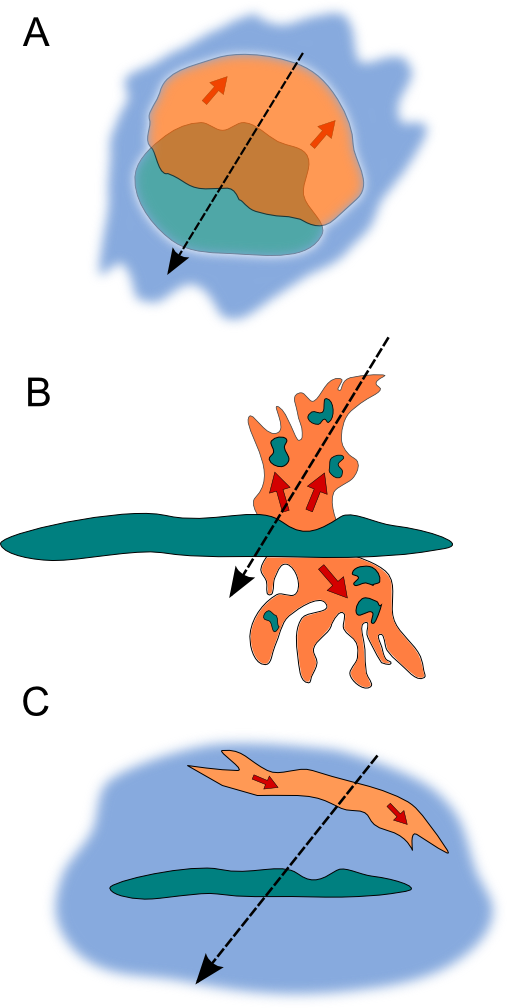}}
\caption {Possible scenarios that could explain the observed absorption 
} \label{scenarios}
\end{center} 
\end{figure}


\subsection{Possible origins of the absorbing gas}

\subsubsection{Stripping (A)}

The weakly ionised absorber at $z=2.304$ might trace a dense cloud,
whose ionised envelope has been stripped away by ram-pressure forces
as the cloud is moving through an ambient (hot) medium.
In this scenario, a cold gas clump (e.g. gas infalling from the IGM)  
moves through hot coronal (circumgalactic) gas that fills the 
potential well of a galaxy (or galaxy group;
see Fig.\ \ref{scenarios}A). The weakly ionised 
absorber would represent the (cold) cloud core, while the redshifted
highly-ionised absorber would represent the former cloud envelope that has been stripped away. The higher metallicity of the stripped material then could be explained by mixing processes of the gas with the ambient 
coronal gas that would need to be metal rich. 

While such processes do play an important role for high-density environments in the cosmic web (such as in galaxy clusters; e.g. \citet{Ebeling2013}), ram pressure stripping in lower-density gas environments such as in galaxy halos is expected to be less efficient, as $p_{\rm ram}
\propto \rho_{\rm gas}$. The most severe drawback of this model
is, however, the very large velocity difference between the cold
cloud core and the stripped envelope of more than $220$\,km\,s$^{-1}$,
which is far beyond any realistic velocity difference expected for 
gas-stripping processes in diffuse gas environments in 
galaxy halos or galaxy groups.
In view of this aspect, the stripping model appears highly 
unlikely to explain the observed properties of the sub-DLA
towards Q\,$0453-423$.

\subsubsection{Galactic outflow (B)}

The weakly ionised absorber at $z=2.304$ might represent a neutral gas disk of a galaxy, in which star-formation activity drives an outflow of ionised gas into the galaxy’s halo.
In this scenario (see Fig.\ \ref{scenarios}C), we think of the central gas region of a galaxy with active star formation. Highly ionised, metal enriched gas might break through the gas layers if supernova explosions provide sufficient mechanical energy, as described in \citet{Fox2007a}. A correlation of intergalactic C\,{\sc iv} absorption and outflows from star-forming galaxies has already been suggested by previous studies, e.g. \citet{Adelberger2005}. If we assume that the sightline passes nearly perpendicular through the neutral disk and along the galactic outflow, we would expect an absorption pattern that mimics the one seen in the absorber pair near $z=2.304$ towards Q\,$0453-423$: a dense and weakly-ionised absorber (gas disk) and highly ionised gas (outflowing material) that has a significant velocity offset due to the large wind velocity.

From our Cloudy modelling, we have estimated an absorber thickness for the weakly ionised absorber of 892\,pc. This relative small number fits well to the values expected for the neutral gas disks of galaxies, in particular if part of the more diffuse neutral gas at the disk boundaries would be removed from the disk by the outflow. Note that with the density and temperature constraints from our Cloudy model the absorbing gas would be regarded as warm neutral medium (WNM). 

Now, that we have identified the neutral phase in this absorber with the inner neutral parts of some form of galaxy, can we identify the highly ionised phase with a galactic wind? 

Previous absorption line studies have measured or inferred outflow speeds of starburst-driven galactic superwinds that range from a few $10^{2}$ to $10^{3}$\,km s$^{-1}$ \citep{Heckman2001}. 
\citet{Steidel2004} find bulk outflow speeds of $200 - 300$\,km s$^{-1}$ to be common for galactic scale winds in star-forming galaxies at $z\leq3$. 
\citet{Bouche2012} find outflow speeds to be 150-300\,km s$^{-1}$ at $z=0$. They trace outflows by Mg\,{\sc ii} absorption and claim that the outflow speeds are lower than the escape velocity of the observed galaxies. 
At intermediate redshifts, \citet{Martin2012} measure outflow speeds at redshifts $0.4-1.4$ to be roughly 200\,km s$^{-1}$, as inferred from low-ion absorption lines.
Also, simulations show that galactic winds can reach velocities of $\sim 500$\,km s$^{-1}$ \citep{Cen2012} and cause multiphase gas layers with distinct temperatures \citep{Cooper2008}. These wind velocities match the velocity difference between the weakly-ionised and highly-ionised absorbers in our spectrum.

The idea of a galactic wind is further supported by the detection of Doppler-shifted O\,{\sc vi} absorption, which is in line with previous findings of \citet{Aracil2004}. These authors show that O\,{\sc vi} absorption is often located within $\sim 300$ km s$^{-1}$ of strong H\,{\sc i} lines, suggesting that the O\,{\sc vi} lines originate in winds flowing away from overdense regions. Also, \citet{Lehner2014} hypothesize that O\,{\sc vi} absorbers at $2 < z \lesssim 3.5$ probe outflows of star-forming galaxies. However, in their study they have exclusively been looking for absorbers with $\log N_{\text{H\,{\sc i}}} > 17.3$. It is especially interesting that for every clear non-detection of O\,{\sc vi}, they do not detect C\,{\sc iv} either.

On the other hand, \citet{Songaila2006} shows that only half of the higher column density C\,{\sc iv} systems have velocity widths that might be produced directly by outflows. She argues that C\,{\sc iv} absorption lines arising in galaxy outflows should show large velocity widths of about $300-400$\,km\,s$^{-1}$. The rest of the higher column density systems and all the lower column density systems must lie in the more general intergalactic medium (see also \citet{Fox2007a}).
However, both narrow and broad C\,{\sc iv} absorption components could trace wind material if the outflowing (and swept-up material from the disk) is clumpy, as indicated by outflowing gas in low-redshift starburst galaxies \citep[e.g.][]{Leroy2015,Roche2015,Bouche2009}. 

Note that in our wind scenario the sightline would pass through hot coronal halo gas on both sides of the disk. The question arises, why there is no C\,{\sc iv} absorption stemming from the disk-halo interface or from outflowing (blueshifted) gas on the other side (near side) of the disk? The answer might again lie in an inhomogeneous density and temperature distribution of the outflowing material such that by chance only the redshifted outflowing material is dense and cold enough to be traced by C\,{\sc iv} and O\,{\sc vi}, while the blueshifted outflowing material along the sightline is too hot and/or too thin to be picked up in absorption.


\subsubsection{Circumgalactic gas flow (C)}

The highly ionised absorber at $z=2.304$ might represent warm/hot gas that resides in the circumgalactic environment of a galaxy whose neutral gas disk is traced by the weakly ionised absorption component.

In this scenario, the highly ionised absorption component may represent a stream of warm/hot gas at larger distances than in the previous scenario, either inside or outside the virial radius of the host galaxy. The velocity offset between the weakly and highly ionised gas would then be a result of the relative motion of the warm/hot gas with respect to the galaxy, for instance as part of an accretion flow from the IGM. The presence of such warm/hot circumgalactic gas flows around galaxies/DLAs/sub-DLAs is well established at low and high redshift \citep[e.g.][]{Simcoe2002,Stocke2006,Prochaska2006,Fox2007b,Lehner2008,Wakker2009}. Also for our Galaxy, FUV observations of O\,{\sc vi} indicate the widespread presence of highly ionised gas features in the inner and outer halo of the Milky Way with velocity separation of several hundred km\,s$^{-1}$ with respect to the disk gas \citep{Sembach2003,Wakker2003}. While the general setup of such a gas distribution (cold disk and warm CGM, kinematically separated by a few hundred km\,s$^{-1})$ is plausible in view of the role of the circumgalactic medium for the growth and evolution of galaxies, the drawback of this scenario is that it alone does not explain the lack of a C\,{\sc iv} absorbing gas phase in the disk-halo interface region. Even more critical is the observed higher metallicity of the highly ionised gas compared to the weakly ionised absorber, which would be difficult to explain if the highly ionised gas is accreted from outside the virial radius of the host galaxy (e.g. from the IGM). We conclude that, in view of these arguments, the circumgalactic gas flow model is less convincing than the outflow model. In summary, the outflow model (B) represents the most realistic scenario that can explain all the observed properties of the absorber pair at $z=2.304$ towards Q\,$0453-423$ based on realistic assumptions.

\subsection{Implications}

A search of the literature reveals that this is not the first absorption system without C\,{\sc iv} absorption that has been found, but it is the first spectrum with this high S/N and which has been analysed in this detail:

\begin{itemize}

\item \citet{Petitjean1994} find an absorption line system with neither Si\,{\sc iv} nor C\,{\sc iv} absorption at a redshift of $z=2.033$ towards PKS 0424-131. Their signal-to-noise is $20-50$. They argue that the sightline passes through a low carbon but enhanced oxygen region.

\item \citet{Ledoux1998} find an absorption system at $z=3.151$ towards Q 2233+131 in which the strongest C\,{\sc iv} absorption is located in a gap between two low-ionisation components.

\item \citet{Petitjean2002} study a damped Lyman $\alpha$ absorber with molecular hydrogen and a complex velocity pattern. Some velocity components are only seen in low ion absorption. It is noteworthy that most of the Si\,{\sc iv} and C\,{\sc iv} absorption is located in between the two main low-ionisation systems.

\item \citet{Peroux2003} find a sub-DLA system where the highly ionised phase does not overlap with the weakly ionised phase, even though the velocity offset is just $\sim 70$\,km\,s$^{-1}$.

\item \citet{Peroux2006} study an extremely metal-rich sub-DLA which shows no third ionisation stages at all. They determine that the gas in this system is predominantly neutral.

\item \citet{Peroux2007} find an absorption system at $z=3.665$ with Si\,{\sc iv} but without C\,{\sc iv} detection  even though the region of the spectrum where the C\,{\sc iv} is expected is noisy. 

\item In a survey of low-ionisation metal absorption systems at $z\sim6$, \citet{Becker2011} discover a lack of strong C\,{\sc iv} and Si\,{\sc iv} absorption. However, their spectra are of modest quality at these far-red wavelengths and the lack of highly ionised metals is still consistent with an overall low metallicity of the systems at these redshifts.

\item In the first results of the KODIAQ survey, \citet{Lehner2014} find three absorption systems without C\,{\sc iv} absorption, even though their S/N is lower than in the spectrum analysed here. Besides, they find 8 systems where the C\,{\sc iv} absorption is offset from the low ion absorption, sometimes by more than $200$\,km s$^{-1}$.

\end{itemize}

The detection of intervening systems without C\,{\sc iv} absorption has not been studied systematically so far. This aspect, however, definitely deserves further attention since such absorbers may provide important constraints on the interpretation of C\,{\sc iv} absorption in the vicinity of optically thick H\,{\sc i} absorbers in quasar spectra.

\section{Summary} 

We have analysed in detail a sub-damped Lyman $\alpha$ absorber at $z=2.304$ towards Q\,0453--423 that shows a variety of weakly ionised metal species, but surprisingly no C\,{\sc iv} absorption in the same velocity range. This is not the first absorption system without C\,{\sc iv} absorption that has been found, but it is the first spectrum with a S/N of up to $\sim100$ that has been analysed in such detail.

A gas phase with highly ionised species is observed at a velocity offset of more than 220\,km\,s$^{-1}$. For both gas phases, we have reproduced the shape of the absorption lines by Voigt profile fitting and analysed the physical conditions in the gas with the help of Cloudy photoionisation models. They show that the gas in the weakly ionised gas phase must be very dense ($n_{\text{H}} > 10^{-1.5}$\,cm$^{-3}$), metalpoor (2.5\% solar), and resemble warm neutral gas in the disk and halo of the Milky Way. By contrast, the gas in the highly ionised phase must be thinner ($n_{\text{H}} \approx 10^{-3.2}$\,cm$^{-3}$) and more metal rich ($\sim17$\% solar). The abundance patterns of the two gas phases seem to be distinct.

We discuss three different scenarios that could explain the observed abundance pattern, taking into consideration that the physical conditions in the weakly ionised phase would result in a very short live span of such a cloud. We conclude that an outflow model, where a galactic wind flows away from a dense region with active star-formation, represents a plausible scenario that can explain the observed properties of the absorber pair.

The detection of absorption line systems without C\,{\sc iv} absorption challenges our interpretation of highly ionised gas in optically thick H\,{\sc i} absorbers in quasar spectra and requires further study.

\begin{acknowledgements}

The authors would like to thank Michael T. Murphy for providing the reduced VLT/UVES data set of Q\,$0453-423$. A.F. is grateful for financial support from the Leibniz Graduate School for Quantitative Spectroscopy in Astrophysics, a joint project of the Leibniz Institute for Astrophysics Potsdam (AIP) and the Institute of Physics and Astronomy of the University of Potsdam (UP). She would also like to thank Andrew Fox, C\'{e}line P\'{e}roux, and Nicolas Lehner 
for helpful discussions. 
 
\end{acknowledgements}

\bibliographystyle{aa}
\bibliography{bibfile}

\begin{thebibliography}{97}
\expandafter\ifx\csname natexlab\endcsname\relax\def\natexlab#1{#1}\fi

\bibitem[{{Adelberger} {et~al.}(2005){Adelberger}, {Shapley}, {Steidel},
  {Pettini}, {Erb}, \& {Reddy}}]{Adelberger2005}
{Adelberger}, K.~L., {Shapley}, A.~E., {Steidel}, C.~C., {et~al.} 2005, \apj,
  629, 636

\bibitem[{{Akerman} {et~al.}(2004){Akerman}, {Carigi}, {Nissen}, {Pettini}, \&
  {Asplund}}]{Akerman2004}
{Akerman}, C.~J., {Carigi}, L., {Nissen}, P.~E., {Pettini}, M., \& {Asplund},
  M. 2004, \aap, 414, 931

\bibitem[{{Aracil} {et~al.}(2004){Aracil}, {Petitjean}, {Pichon}, \&
  {Bergeron}}]{Aracil2004}
{Aracil}, B., {Petitjean}, P., {Pichon}, C., \& {Bergeron}, J. 2004, \aap, 419,
  811

\bibitem[{{Asplund} {et~al.}(2009){Asplund}, {Grevesse}, {Sauval}, \&
  {Scott}}]{Asplund2009}
{Asplund}, M., {Grevesse}, N., {Sauval}, A.~J., \& {Scott}, P. 2009, \araa, 47,
  481

\bibitem[{{Becker} {et~al.}(2011){Becker}, {Sargent}, {Rauch}, \&
  {Calverley}}]{Becker2011}
{Becker}, G.~D., {Sargent}, W.~L.~W., {Rauch}, M., \& {Calverley}, A.~P. 2011,
  \apj, 735, 93

\bibitem[{{Bergeron} {et~al.}(1994){Bergeron}, {Petitjean}, {Sargent},
  {Bahcall}, {Boksenberg}, {Hartig}, {Jannuzi}, {Kirhakos}, {Savage},
  {Schneider}, {Turnshek}, {Weymann}, \& {Wolfe}}]{Bergeron1994}
{Bergeron}, J., {Petitjean}, P., {Sargent}, W.~L.~W., {et~al.} 1994, \apj, 436,
  33

\bibitem[{{Birnboim} \& {Dekel}(2003)}]{Birnboim2003}
{Birnboim}, Y. \& {Dekel}, A. 2003, \mnras, 345, 349

\bibitem[{{Boksenberg} {et~al.}(2003){Boksenberg}, {Sargent}, \&
  {Rauch}}]{Boksenberg2003}
{Boksenberg}, A., {Sargent}, W.~L.~W., \& {Rauch}, M. 2003, ArXiv Astrophysics
  e-prints

\bibitem[{{Bouch{\'e}} {et~al.}(2012){Bouch{\'e}}, {Hohensee}, {Vargas},
  {Kacprzak}, {Martin}, {Cooke}, \& {Churchill}}]{Bouche2012}
{Bouch{\'e}}, N., {Hohensee}, W., {Vargas}, R., {et~al.} 2012, \mnras, 426, 801

\bibitem[{{Bouch{\'e}} {et~al.}(2007){Bouch{\'e}}, {Lehnert}, {Aguirre},
  {P{\'e}roux}, \& {Bergeron}}]{Bouche2007}
{Bouch{\'e}}, N., {Lehnert}, M.~D., {Aguirre}, A., {P{\'e}roux}, C., \&
  {Bergeron}, J. 2007, \mnras, 378, 525

\bibitem[{{Bouch{\'e}} {et~al.}(2005){Bouch{\'e}}, {Lehnert}, \&
  {P{\'e}roux}}]{Bouche2005}
{Bouch{\'e}}, N., {Lehnert}, M.~D., \& {P{\'e}roux}, C. 2005, \mnras, 364, 319

\bibitem[{{Bouch{\'e}} {et~al.}(2006){Bouch{\'e}}, {Lehnert}, \&
  {P{\'e}roux}}]{Bouche2006}
{Bouch{\'e}}, N., {Lehnert}, M.~D., \& {P{\'e}roux}, C. 2006, \mnras, 367, L16

\bibitem[{{Cen}(2012)}]{Cen2012}
{Cen}, R. 2012, \apj, 748, 121

\bibitem[{{Cooke} {et~al.}(2011){Cooke}, {Pettini}, {Steidel}, {Rudie}, \&
  {Nissen}}]{Cooke2011}
{Cooke}, R., {Pettini}, M., {Steidel}, C.~C., {Rudie}, G.~C., \& {Nissen},
  P.~E. 2011, \mnras, 417, 1534

\bibitem[{{Cooper} {et~al.}(2008){Cooper}, {Bicknell}, {Sutherland}, \&
  {Bland-Hawthorn}}]{Cooper2008}
{Cooper}, J.~L., {Bicknell}, G.~V., {Sutherland}, R.~S., \& {Bland-Hawthorn},
  J. 2008, \apj, 674, 157

\bibitem[{{Cowie} \& {Songaila}(1998)}]{Cowie1998Nature}
{Cowie}, L.~L. \& {Songaila}, A. 1998, \nat, 394, 44

\bibitem[{{Crighton} {et~al.}(2015){Crighton}, {Hennawi}, {Simcoe}, {Cooksey},
  {Murphy}, {Fumagalli}, {Prochaska}, \& {Shanks}}]{Crighton2015}
{Crighton}, N.~H.~M., {Hennawi}, J.~F., {Simcoe}, R.~A., {et~al.} 2015, \mnras,
  446, 18

\bibitem[{{Dav{\'e}} {et~al.}(2001){Dav{\'e}}, {Cen}, {Ostriker}, {Bryan},
  {Hernquist}, {Katz}, {Weinberg}, {Norman}, \& {O'Shea}}]{Dave2001}
{Dav{\'e}}, R., {Cen}, R., {Ostriker}, J.~P., {et~al.} 2001, \apj, 552, 473

\bibitem[{{Dessauges-Zavadsky} {et~al.}(2003){Dessauges-Zavadsky},
  {P{\'e}roux}, {Kim}, {D'Odorico}, \& {McMahon}}]{Dessauges2003}
{Dessauges-Zavadsky}, M., {P{\'e}roux}, C., {Kim}, T.-S., {D'Odorico}, S., \&
  {McMahon}, R.~G. 2003, \mnras, 345, 447

\bibitem[{{D'Odorico} {et~al.}(2010){D'Odorico}, {Calura}, {Cristiani}, \&
  {Viel}}]{D'Odorico2010}
{D'Odorico}, V., {Calura}, F., {Cristiani}, S., \& {Viel}, M. 2010, \mnras,
  401, 2715

\bibitem[{{D'Odorico} {et~al.}(2013){D'Odorico}, {Cupani}, {Cristiani},
  {Maiolino}, {Molaro}, {Nonino}, {Centuri{\'o}n}, {Cimatti}, {di Serego
  Alighieri}, {Fiore}, {Fontana}, {Gallerani}, {Giallongo}, {Mannucci},
  {Marconi}, {Pentericci}, {Viel}, \& {Vladilo}}]{D'Odorico2013}
{D'Odorico}, V., {Cupani}, G., {Cristiani}, S., {et~al.} 2013, \mnras, 435,
  1198

\bibitem[{{Ebeling} {et~al.}(2013){Ebeling}, {Edge}, {Burgett}, {Chambers},
  {Hodapp}, {Huber}, {Kaiser}, {Price}, \& {Tonry}}]{Ebeling2013}
{Ebeling}, H., {Edge}, A.~C., {Burgett}, W.~S., {et~al.} 2013, \mnras, 432, 62

\bibitem[{{Efstathiou}(2000)}]{Efstathiou2000}
{Efstathiou}, G. 2000, \mnras, 317, 697

\bibitem[{{Ellison} {et~al.}(2000){Ellison}, {Songaila}, {Schaye}, \&
  {Pettini}}]{Ellison2000}
{Ellison}, S.~L., {Songaila}, A., {Schaye}, J., \& {Pettini}, M. 2000, \aj,
  120, 1175

\bibitem[{{Erb} {et~al.}(2006){Erb}, {Shapley}, {Pettini}, {Steidel}, {Reddy},
  \& {Adelberger}}]{Erb2006}
{Erb}, D.~K., {Shapley}, A.~E., {Pettini}, M., {et~al.} 2006, \apj, 644, 813

\bibitem[{{Fangano} {et~al.}(2007){Fangano}, {Ferrara}, \&
  {Richter}}]{Fangano2007}
{Fangano}, A.~P.~M., {Ferrara}, A., \& {Richter}, P. 2007, \mnras, 381, 469

\bibitem[{{Ferland} {et~al.}(1998){Ferland}, {Korista}, {Verner}, {Ferguson},
  {Kingdon}, \& {Verner}}]{Ferland1998}
{Ferland}, G.~J., {Korista}, K.~T., {Verner}, D.~A., {et~al.} 1998, \pasp, 110,
  761

\bibitem[{{Ferrara} {et~al.}(2005){Ferrara}, {Scannapieco}, \&
  {Bergeron}}]{Ferrara2005}
{Ferrara}, A., {Scannapieco}, E., \& {Bergeron}, J. 2005, \apjl, 634, L37

\bibitem[{{Fontana} \& {Ballester}(1995)}]{Fontana1995}
{Fontana}, A. \& {Ballester}, P. 1995, The Messenger, 80, 37

\bibitem[{{Fox} {et~al.}(2014){Fox}, {Richter}, \& {Fechner}}]{meinPaper2014}
{Fox}, A., {Richter}, P., \& {Fechner}, C. 2014, \aap, 572, A102

\bibitem[{{Fox}(2011)}]{Fox2011a}
{Fox}, A.~J. 2011, \apj, 730, 58

\bibitem[{{Fox} {et~al.}(2007{\natexlab{a}}){Fox}, {Ledoux}, {Petitjean}, \&
  {Srianand}}]{Fox2007a}
{Fox}, A.~J., {Ledoux}, C., {Petitjean}, P., \& {Srianand}, R.
  2007{\natexlab{a}}, \aap, 473, 791

\bibitem[{{Fox} {et~al.}(2011){Fox}, {Ledoux}, {Petitjean}, {Srianand}, \&
  {Guimar{\~a}es}}]{Fox2011}
{Fox}, A.~J., {Ledoux}, C., {Petitjean}, P., {Srianand}, R., \&
  {Guimar{\~a}es}, R. 2011, \aap, 534, A82

\bibitem[{{Fox} {et~al.}(2007{\natexlab{b}}){Fox}, {Petitjean}, {Ledoux}, \&
  {Srianand}}]{Fox2007b}
{Fox}, A.~J., {Petitjean}, P., {Ledoux}, C., \& {Srianand}, R.
  2007{\natexlab{b}}, \aap, 465, 171

\bibitem[{{Fox} {et~al.}(2007{\natexlab{c}}){Fox}, {Petitjean}, {Ledoux}, \&
  {Srianand}}]{Fox2007c}
{Fox}, A.~J., {Petitjean}, P., {Ledoux}, C., \& {Srianand}, R.
  2007{\natexlab{c}}, \apjl, 668, L15

\bibitem[{{Gnedin} \& {Ostriker}(1997)}]{Gnedin1997}
{Gnedin}, N.~Y. \& {Ostriker}, J.~P. 1997, \apj, 486, 581

\bibitem[{{Haardt} \& {Madau}(2012)}]{Haardt2012}
{Haardt}, F. \& {Madau}, P. 2012, \apj, 746, 125

\bibitem[{{Heckman}(2001)}]{Heckman2001}
{Heckman}, T.~M. 2001, in Astronomical Society of the Pacific Conference
  Series, Vol. 240, Gas and Galaxy Evolution, ed. J.~E. {Hibbard}, M.~{Rupen},
  \& J.~H. {van Gorkom}, 345

\bibitem[{{Heckman} {et~al.}(2002){Heckman}, {Norman}, {Strickland}, \&
  {Sembach}}]{Heckman2002}
{Heckman}, T.~M., {Norman}, C.~A., {Strickland}, D.~K., \& {Sembach}, K.~R.
  2002, \apj, 577, 691

\bibitem[{{Kang} {et~al.}(2005){Kang}, {Ryu}, {Cen}, \& {Song}}]{Kang2005}
{Kang}, H., {Ryu}, D., {Cen}, R., \& {Song}, D. 2005, \apj, 620, 21

\bibitem[{{Kawata} \& {Rauch}(2007)}]{Kawata2007}
{Kawata}, D. \& {Rauch}, M. 2007, \apj, 663, 38

\bibitem[{{Kere{\v s}} {et~al.}(2005){Kere{\v s}}, {Katz}, {Weinberg}, \&
  {Dav{\'e}}}]{Keres2005}
{Kere{\v s}}, D., {Katz}, N., {Weinberg}, D.~H., \& {Dav{\'e}}, R. 2005,
  \mnras, 363, 2

\bibitem[{{Kirkman} \& {Tytler}(1999)}]{Kirkman1999}
{Kirkman}, D. \& {Tytler}, D. 1999, \apjl, 512, L5

\bibitem[{{Kwak} \& {Shelton}(2010)}]{Kwak2010}
{Kwak}, K. \& {Shelton}, R.~L. 2010, \apj, 719, 523

\bibitem[{{Ledoux} {et~al.}(1998){Ledoux}, {Petitjean}, {Bergeron}, {Wampler},
  \& {Srianand}}]{Ledoux1998}
{Ledoux}, C., {Petitjean}, P., {Bergeron}, J., {Wampler}, E.~J., \& {Srianand},
  R. 1998, \aap, 337, 51

\bibitem[{{Lehner} {et~al.}(2008){Lehner}, {Howk}, {Prochaska}, \&
  {Wolfe}}]{Lehner2008}
{Lehner}, N., {Howk}, J.~C., {Prochaska}, J.~X., \& {Wolfe}, A.~M. 2008,
  \mnras, 390, 2

\bibitem[{{Lehner} {et~al.}(2014){Lehner}, {O'Meara}, {Fox}, {Howk},
  {Prochaska}, {Burns}, \& {Armstrong}}]{Lehner2014}
{Lehner}, N., {O'Meara}, J.~M., {Fox}, A.~J., {et~al.} 2014, \apj, 788, 119

\bibitem[{{Leroy} {et~al.}(2015){Leroy}, {Walter}, {Martini}, {Roussel},
  {Sandstrom}, {Ott}, {Weiss}, {Bolatto}, {Schuster}, \&
  {Dessauges-Zavadsky}}]{Leroy2015}
{Leroy}, A.~K., {Walter}, F., {Martini}, P., {et~al.} 2015, \apj, 814, 83

\bibitem[{{Lu} {et~al.}(1996){Lu}, {Sargent}, {Barlow}, {Churchill}, \&
  {Vogt}}]{Lu1996}
{Lu}, L., {Sargent}, W.~L.~W., {Barlow}, T.~A., {Churchill}, C.~W., \& {Vogt},
  S.~S. 1996, \apjs, 107, 475

\bibitem[{{Madau} {et~al.}(2001){Madau}, {Ferrara}, \& {Rees}}]{Madau2001}
{Madau}, P., {Ferrara}, A., \& {Rees}, M.~J. 2001, \apj, 555, 92

\bibitem[{{Martin} \& {Bouch{\'e}}(2009)}]{Bouche2009}
{Martin}, C.~L. \& {Bouch{\'e}}, N. 2009, \apj, 703, 1394

\bibitem[{{Martin} {et~al.}(2010){Martin}, {Scannapieco}, {Ellison}, {Hennawi},
  {Djorgovski}, \& {Fournier}}]{Martin2010}
{Martin}, C.~L., {Scannapieco}, E., {Ellison}, S.~L., {et~al.} 2010, \apj, 721,
  174

\bibitem[{{Martin} {et~al.}(2012){Martin}, {Shapley}, {Coil}, {Kornei},
  {Bundy}, {Weiner}, {Noeske}, \& {Schiminovich}}]{Martin2012}
{Martin}, C.~L., {Shapley}, A.~E., {Coil}, A.~L., {et~al.} 2012, \apj, 760, 127

\bibitem[{{Meiksin}(2009)}]{Meiksin2009}
{Meiksin}, A.~A. 2009, Reviews of Modern Physics, 81, 1405

\bibitem[{{Morton}(2003)}]{Morton2003}
{Morton}, D.~C. 2003, \apjs, 149, 205

\bibitem[{{Nagamine} {et~al.}(2010){Nagamine}, {Choi}, \&
  {Yajima}}]{Nagamine2010}
{Nagamine}, K., {Choi}, J.-H., \& {Yajima}, H. 2010, \apjl, 725, L219

\bibitem[{{Oppenheimer} \& {Dav{\'e}}(2006)}]{Oppenheimer2006}
{Oppenheimer}, B.~D. \& {Dav{\'e}}, R. 2006, \mnras, 373, 1265

\bibitem[{{Pagel}(1999)}]{Pagel1999}
{Pagel}, B.~E.~J. 1999, ArXiv Astrophysics e-prints

\bibitem[{{P{\'e}roux} {et~al.}(2003){P{\'e}roux}, {Dessauges-Zavadsky},
  {D'Odorico}, {Kim}, \& {McMahon}}]{Peroux2003}
{P{\'e}roux}, C., {Dessauges-Zavadsky}, M., {D'Odorico}, S., {Kim}, T.-S., \&
  {McMahon}, R.~G. 2003, \mnras, 345, 480

\bibitem[{{P{\'e}roux} {et~al.}(2007){P{\'e}roux}, {Dessauges-Zavadsky},
  {D'Odorico}, {Kim}, \& {McMahon}}]{Peroux2007}
{P{\'e}roux}, C., {Dessauges-Zavadsky}, M., {D'Odorico}, S., {Kim}, T.-S., \&
  {McMahon}, R.~G. 2007, \mnras, 382, 177

\bibitem[{{P{\'e}roux} {et~al.}(2006){P{\'e}roux}, {Kulkarni}, {Meiring},
  {Ferlet}, {Khare}, {Lauroesch}, {Vladilo}, \& {York}}]{Peroux2006}
{P{\'e}roux}, C., {Kulkarni}, V.~P., {Meiring}, J., {et~al.} 2006, \aap, 450,
  53

\bibitem[{{Petitjean} {et~al.}(1994){Petitjean}, {Rauch}, \&
  {Carswell}}]{Petitjean1994}
{Petitjean}, P., {Rauch}, M., \& {Carswell}, R.~F. 1994, \aap, 291, 29

\bibitem[{{Petitjean} {et~al.}(2002){Petitjean}, {Srianand}, \&
  {Ledoux}}]{Petitjean2002}
{Petitjean}, P., {Srianand}, R., \& {Ledoux}, C. 2002, \mnras, 332, 383

\bibitem[{{Pettini}(1999)}]{Pettini1999}
{Pettini}, M. 1999, in Chemical Evolution from Zero to High Redshift, ed. J.~R.
  {Walsh} \& M.~R. {Rosa}, 233

\bibitem[{{Pettini} {et~al.}(2003){Pettini}, {Madau}, {Bolte}, {Prochaska},
  {Ellison}, \& {Fan}}]{Pettini2003}
{Pettini}, M., {Madau}, P., {Bolte}, M., {et~al.} 2003, \apj, 594, 695

\bibitem[{{Pettini} {et~al.}(2001){Pettini}, {Shapley}, {Steidel}, {Cuby},
  {Dickinson}, {Moorwood}, {Adelberger}, \& {Giavalisco}}]{Pettini2001}
{Pettini}, M., {Shapley}, A.~E., {Steidel}, C.~C., {et~al.} 2001, \apj, 554,
  981

\bibitem[{{Pettini} {et~al.}(2008){Pettini}, {Zych}, {Steidel}, \&
  {Chaffee}}]{Pettini2008}
{Pettini}, M., {Zych}, B.~J., {Steidel}, C.~C., \& {Chaffee}, F.~H. 2008,
  \mnras, 385, 2011

\bibitem[{{Prochaska} {et~al.}(2006){Prochaska}, {Weiner}, {Chen}, \&
  {Mulchaey}}]{Prochaska2006}
{Prochaska}, J.~X., {Weiner}, B.~J., {Chen}, H.-W., \& {Mulchaey}, J.~S. 2006,
  \apj, 643, 680

\bibitem[{{Prochter} {et~al.}(2010){Prochter}, {Prochaska}, {O'Meara},
  {Burles}, \& {Bernstein}}]{Prochter2010}
{Prochter}, G.~E., {Prochaska}, J.~X., {O'Meara}, J.~M., {Burles}, S., \&
  {Bernstein}, R.~A. 2010, \apj, 708, 1221

\bibitem[{{Rafelski} {et~al.}(2012){Rafelski}, {Wolfe}, {Prochaska},
  {Neeleman}, \& {Mendez}}]{Rafelski2012}
{Rafelski}, M., {Wolfe}, A.~M., {Prochaska}, J.~X., {Neeleman}, M., \&
  {Mendez}, A.~J. 2012, \apj, 755, 89

\bibitem[{{Rahmati} {et~al.}(2013){Rahmati}, {Pawlik}, {Raicevic}, \&
  {Schaye}}]{Rahmati2013}
{Rahmati}, A., {Pawlik}, A.~H., {Raicevic}, M., \& {Schaye}, J. 2013, \mnras,
  430, 2427

\bibitem[{{Rees} \& {Ostriker}(1977)}]{ReesOstriker1977}
{Rees}, M.~J. \& {Ostriker}, J.~P. 1977, \mnras, 179, 541

\bibitem[{{Richter} {et~al.}(2009){Richter}, {Charlton}, {Fangano}, {Bekhti},
  \& {Masiero}}]{Richter2009}
{Richter}, P., {Charlton}, J.~C., {Fangano}, A.~P.~M., {Bekhti}, N.~B., \&
  {Masiero}, J.~R. 2009, \apj, 695, 1631

\bibitem[{{Richter} {et~al.}(2005){Richter}, {Ledoux}, {Petitjean}, \&
  {Bergeron}}]{Richter2005}
{Richter}, P., {Ledoux}, C., {Petitjean}, P., \& {Bergeron}, J. 2005, \aap,
  440, 819

\bibitem[{{Roche} {et~al.}(2015){Roche}, {Humphrey}, {Gomes}, {Papaderos},
  {Lagos}, \& {S{\'a}nchez}}]{Roche2015}
{Roche}, N., {Humphrey}, A., {Gomes}, J.~M., {et~al.} 2015, \mnras, 453, 2349

\bibitem[{{Ryan-Weber} {et~al.}(2006){Ryan-Weber}, {Pettini}, \&
  {Madau}}]{Ryan-Weber2006}
{Ryan-Weber}, E.~V., {Pettini}, M., \& {Madau}, P. 2006, \mnras, 371, L78

\bibitem[{{Ryan-Weber} {et~al.}(2009){Ryan-Weber}, {Pettini}, {Madau}, \&
  {Zych}}]{Ryan-Weber2009}
{Ryan-Weber}, E.~V., {Pettini}, M., {Madau}, P., \& {Zych}, B.~J. 2009, \mnras,
  395, 1476

\bibitem[{{Schaye}(2001)}]{Schaye2001}
{Schaye}, J. 2001, \apjl, 562, L95

\bibitem[{{Schaye} {et~al.}(2003){Schaye}, {Aguirre}, {Kim}, {Theuns}, {Rauch},
  \& {Sargent}}]{Schaye2003}
{Schaye}, J., {Aguirre}, A., {Kim}, T.-S., {et~al.} 2003, \apj, 596, 768

\bibitem[{{Schaye} {et~al.}(2007){Schaye}, {Carswell}, \& {Kim}}]{Schaye2007}
{Schaye}, J., {Carswell}, R.~F., \& {Kim}, T.-S. 2007, \mnras, 379, 1169

\bibitem[{{Sembach} {et~al.}(2003){Sembach}, {Wakker}, {Savage}, {Richter},
  {Meade}, {Shull}, {Jenkins}, {Sonneborn}, \& {Moos}}]{Sembach2003}
{Sembach}, K.~R., {Wakker}, B.~P., {Savage}, B.~D., {et~al.} 2003, \apjs, 146,
  165

\bibitem[{{Simcoe}(2006)}]{Simcoe2006}
{Simcoe}, R.~A. 2006, \apj, 653, 977

\bibitem[{{Simcoe} {et~al.}(2002){Simcoe}, {Sargent}, \& {Rauch}}]{Simcoe2002}
{Simcoe}, R.~A., {Sargent}, W.~L.~W., \& {Rauch}, M. 2002, \apj, 578, 737

\bibitem[{{Som} {et~al.}(2013){Som}, {Kulkarni}, {Meiring}, {York},
  {P{\'e}roux}, {Khare}, \& {Lauroesch}}]{Som2013}
{Som}, D., {Kulkarni}, V.~P., {Meiring}, J., {et~al.} 2013, \mnras, 435, 1469

\bibitem[{{Sommer-Larsen} \& {Fynbo}(2008)}]{Sommer-Larsen2008}
{Sommer-Larsen}, J. \& {Fynbo}, J.~P.~U. 2008, \mnras, 385, 3

\bibitem[{{Songaila}(2001)}]{Songaila2001}
{Songaila}, A. 2001, \apjl, 561, L153

\bibitem[{{Songaila}(2005)}]{Songaila2005}
{Songaila}, A. 2005, \aj, 130, 1996

\bibitem[{{Songaila}(2006)}]{Songaila2006}
{Songaila}, A. 2006, \aj, 131, 24

\bibitem[{{Steidel} {et~al.}(2001){Steidel}, {Pettini}, \&
  {Adelberger}}]{Steidel2001}
{Steidel}, C.~C., {Pettini}, M., \& {Adelberger}, K.~L. 2001, \apj, 546, 665

\bibitem[{{Steidel} {et~al.}(2004){Steidel}, {Shapley}, {Pettini},
  {Adelberger}, {Erb}, {Reddy}, \& {Hunt}}]{Steidel2004}
{Steidel}, C.~C., {Shapley}, A.~E., {Pettini}, M., {et~al.} 2004, \apj, 604,
  534

\bibitem[{{Stocke} {et~al.}(2006){Stocke}, {Penton}, {Danforth}, {Shull},
  {Tumlinson}, \& {McLin}}]{Stocke2006}
{Stocke}, J.~T., {Penton}, S.~V., {Danforth}, C.~W., {et~al.} 2006, \apj, 641,
  217

\bibitem[{{Veilleux} {et~al.}(2005){Veilleux}, {Cecil}, \&
  {Bland-Hawthorn}}]{Veilleux2005}
{Veilleux}, S., {Cecil}, G., \& {Bland-Hawthorn}, J. 2005, \araa, 43, 769

\bibitem[{{Wakker} \& {Savage}(2009)}]{Wakker2009}
{Wakker}, B.~P. \& {Savage}, B.~D. 2009, \apjs, 182, 378

\bibitem[{{Wakker} {et~al.}(2003){Wakker}, {Savage}, {Sembach}, {Richter},
  {Meade}, {Jenkins}, {Shull}, {Ake}, {Blair}, {Dixon}, {Friedman}, {Green},
  {Green}, {Kruk}, {Moos}, {Murphy}, {Oegerle}, {Sahnow}, {Sonneborn},
  {Wilkinson}, \& {York}}]{Wakker2003}
{Wakker}, B.~P., {Savage}, B.~D., {Sembach}, K.~R., {et~al.} 2003, \apjs, 146,
  1

\bibitem[{{Wolfe} {et~al.}(2005){Wolfe}, {Gawiser}, \& {Prochaska}}]{Wolfe2005}
{Wolfe}, A.~M., {Gawiser}, E., \& {Prochaska}, J.~X. 2005, \araa, 43, 861

\bibitem[{{Wolfe} \& {Prochaska}(2000)}]{Wolfe2000}
{Wolfe}, A.~M. \& {Prochaska}, J.~X. 2000, \apj, 545, 591

\bibitem[{{Yajima} {et~al.}(2012){Yajima}, {Choi}, \& {Nagamine}}]{Yajima2012}
{Yajima}, H., {Choi}, J.-H., \& {Nagamine}, K. 2012, \mnras, 427, 2889

\end{thebibliography}




\end{document}